\documentclass[useAMS,usenatbib]{mn2e}
\newcommand{\xmmn}{\mbox{\em XMM-Newton}}

\newcommand{\asca}{\mbox{\em ASCA}}
\newcommand{\exosat}{\mbox{\em EXOSAT}}

\newcommand{\ginga}{\mbox{\em Ginga}}
\newcommand{\etal}{\mbox{et\ al.\ }}

\newcommand{\msun}{\,\mbox{$\mbox{M}_{\odot}$}}

\newcommand{\aanda}  {A\&A\nolinebreak }

\newcommand{\mn}     {MNRAS\nolinebreak }
\usepackage{rotating}

\title[The X-ray spectrum of NGC 7213 and the Seyfert--LINER connection]
{The X-ray spectrum of NGC 7213 and the Seyfert--LINER connection}

\author[R.L.C. Starling \& M.J. Page \etal]
{R. L. C. Starling$^{1,2}$\thanks{E-mail: starling@science.uva.nl}, M. J. Page$^1$, G. Branduardi-Raymont$^1$, A. A. Breeveld$^1$, \and R. Soria$^1$ $\&$ K. Wu$^1$\\
$^1$Mullard Space Science Laboratory, University College London, Holmbury
St. Mary, Dorking, Surrey RH5 6NT, UK \\
$^2$Astronomical Institute, University of Amsterdam, Kruislaan 403, 1098 SJ Amsterdam, The Netherlands}

\begin{document}
\date{Accepted . Received ; in original form }
%\bigskip

\pagerange{\pageref{firstpage}--\pageref{lastpage}} \pubyear{2004}

\maketitle

\label{firstpage}

%%***************************************************************************
%%  ABSTRACT
%%***************************************************************************

\begin{abstract}
We present an \xmmn\ observation of the Seyfert-LINER galaxy NGC
7213. 
The RGS soft X-ray spectrum is well fitted with a power law plus soft X-ray collisionally ionised thermal plasma ($kT = 0.18^{+0.03}_{-0.01}$ keV). We confirm the presence of Fe I, XXV and XXVI K$\alpha$ emission in the EPIC spectrum and set tighter constraints on their equivalent widths of 82$^{+10}_{-13}$, 24$^{+9}_{-11}$ and 24$^{+10}_{-13}$ eV respectively. 
We compare the observed properties together with the inferred mass accretion
rate of NGC~7213, to those of other
Seyfert and LINER galaxies. We find that NGC~7213 has intermediate 
X-ray spectral properties lying between those of the weak AGN found
in the LINER M\,81 and higher luminosity Seyfert galaxies.
There appears to be a continuous sequence of X-ray properties from the
Galactic Centre through LINER galaxies to Seyferts, likely 
determined by the amount of material available for accretion
in the central regions.
\end{abstract}


\section{Introduction}
Low ionisation nuclear emission line region (LINER) galaxies are  
characterised by optical emission line ratios which indicate
a low level of ionisation \citep{Heckman}. 
The origin of 
these emission lines is still
the subject of debate: the lines are attributed either to shock heating (Baldwin, Phillips \& Terlevich 1981) or to photoionisation
by a central AGN (Ferland \& Netzer 1983; Halpern \& Steiner 1983). 
For example, Dopita \& Sutherland (1996) have shown that shock heating can produce the
optical line
ratios observed in LINER galaxies, with shock velocities of 150 - 500 km
s$^{-1}$. 
Such shocks can originate from stellar processes (e.g. supernovae), and 
can also be produced as a result of a jet or a wind from an AGN.
On the other hand, when the optical line ratios of LINERs were compared to 
those measured
in Seyfert galaxies, Ferland \& Netzer (1983) found that they could explain the
lines in both types of object using the same photoionising AGN 
continuum shape, but 
with a systematic
variation of ionisation parameter,
from low values in LINERs to high values
in Seyferts. 

NGC 7213 is a nearby ($z$ = 0.005977) S0 galaxy with AGN and LINER 
characteristics. It is clear 
that there is an AGN in this source, classified as a
Seyfert 1 from its H$\alpha$ line width (full width at zero intensity $\sim$ 13000 km s$^{-1}$,
Phillips 1979).
A variety of optical emission lines are observed in this galaxy with
velocities ranging from 200 to 2000 km s$^{-1}$ FWHM (Filippenko \& Halpern
1984, hereafter FH84).
FH84 argue that photoionisation of clouds spanning a range of densities and velocities by a
non-stellar continuum is likely to be the mechanism creating the optical line emission. By invoking high density
rather than high temperature clouds, FH84 eliminate the need for shock
heating.

Since its discovery as a low luminosity X-ray source (Marshall \etal 1978)
NGC 7213 has
been observed 
with several X-ray missions.
The presence of a soft X-ray excess in NGC~7213 was implied by
the results of an \exosat\ spectral survey of AGN \citep{exosat} when the measured
absorbing column for a single power law fit was found to be significantly lower than the
Galactic value. In addition, the UV flux measured by Wu, Boggess \& Gull (1983)
was higher than would be expected from an extrapolation of the optical
flux indicating that NGC~7213 may have a big blue
bump (BBB), although weak compared to most Seyfert galaxies. The BBB
is often interpreted as thermal emission from 
an accretion disc; if accretion discs are present in LINERs it is important to determine their properties if we are to understand the underlying
emission mechanisms. 

Turner \& Pounds (1989) and Pounds \etal (1994) reported possible
Fe K$\alpha$ emission in the X-ray spectrum observed with \exosat\ and \ginga, but there was not sufficient resolution to
unambiguously detect these features. 
Recently Bianchi \etal (2003) combined
data from the pn camera on \xmmn\ with simultaneous data from the PDS instrument on
\emph{BeppoSAX} to investigate the iron line complex and reflection hump.
They found that a neutral 
Fe K$\alpha$ line is present, with excess emission at higher energies best explained as weak narrow emission lines from highly ionised iron. The data do not show the presence of a significant reflection hump, suggesting that the neutral Fe K$\alpha$ line originates in Compton-thin material. They also suggest that the highly ionised Fe emission may arise in material photoionised by the AGN power law continuum. 

Here we present 
an analysis of the high resolution RGS spectra, 0.3 - 2 keV EPIC spectra and OM photometry taken during the same \xmmn\ pointing. We revisit the iron line complex with an analysis of the combined pn and MOS1 spectrum in which it is possible to obtain better constraints on individual line features.
Identifying the physical mechanisms producing the X-ray
emission may provide some clues to the origin of the optical emission lines
where at present neither shock heating nor photoionisation by the AGN can be
ruled out. We also compare the X-ray spectral properties of NGC~7213 with
those of the nearest LINER galaxy M\,81, which has also been
studied in detail using
\xmmn\ (Page \etal 2003, 2004), and discuss the relationship between Seyfert galaxies and LINERs. 

\section{XMM-Newton Observation}
NGC 7213 was observed on 2001 May 28/29 with \xmmn\ \citep{XMM} in the RGS Guaranteed Time Programme. The EPIC (Str\"{u}der \etal 2001; Turner \etal 2001) MOS1 and pn cameras were operated in small window mode
with the medium filter. The EPIC MOS2 was in full frame mode (also
medium filter) to image the entire galaxy. MOS2 will be omitted from the analysis due to pile-up, since the count
rate was over three times higher than the recommended limit of 0.7 counts s$^{-1}$ for
point sources \citep{XMMHB}. The RGS instruments \citep{RGS} were in
standard spectroscopy mode. The exposure times are 46448 s for MOS1, 42201 s for pn and 
46716 s for each RGS instrument. No
variability in flux was found during the observation so we produced a single
spectrum from the entire observation for each X-ray instrument. The Optical Monitor was operated in imaging mode with optical and ultraviolet filters.

The RGS data were initially processed using the \xmmn\ SAS
(Science Analysis Software) version
5.2. In each RGS, first and second orders were selected using
pulse-height/dispersion regions containing 93 per cent of the pulse height
distribution. Source+background spectra were taken from spatial regions
containing 90 per cent of the point-spread function in the cross dispersion
direction. Background spectra were obtained from regions excluding 95 per cent of the
point-spread function. Response matrices were generated using the standard SAS
rgsrmfgen task. The effective area of each response matrix was then divided by the ratio
of a power law plus Galactic column fit to the \xmmn\ rev 0084 RGS spectra of the
continuum source Mrk 421 to correct for the residual artifacts in the effective
area calibration. After background subtraction, the individual first and
second order spectra from RGS1 and RGS2 were resampled and coadded to produce a
single spectrum, and the response matrices were combined accordingly.

The raw EPIC data were processed with the \xmmn\ SAS
version 5.4. Spectra were constructed using single and
double events in the pn and all valid event patterns (0-12) in MOS1. Source
counts were taken from a circular region of radius 45\arcsec. In MOS1 a
background spectrum was obtained from a nearby source-free region, 7 times larger in radius
than the source extraction region. For the pn the background spectrum was taken
from 3 separate circular regions
totalling 5.76 times the area of the source circle. The
MOS1 and pn spectra 
were then combined into a single spectrum with 45 eV bins using the method of
Page, Davis \& Salvi (2003) to improve the signal to noise. The March 2003 
EPIC response files for small window were used with the resulting spectra. 

The OM \citep{OM} data were processed with \xmmn\ SAS version 5.4. For each of the
three UV filters the sub-exposures were co-added and corrected
for modulo-8 noise. The total exposure times amount to 4000 s, 7200 s and
8000 s for the UVW1, UVM2 and UVW2 filters respectively. Aperture 
photometry was performed using standard
routines from IRAF. To minimise the contribution of the host galaxy a
circular aperture of 4\arcsec\ in diameter was used to measure the AGN flux,
with a background determined from an annulus of inner diameter 5\arcsec\
and outer diameter 7\arcsec. The resultant magnitudes were then corrected 
for the fraction of the point-spread function falling outside the aperture,
deadtime, coincidence loss, and Galactic reddening (using the reddening law of
Seaton 1979). 

\section{Results}
Spectral analyses of the data were 
performed using the XSPEC v11.2 X-ray spectral
fitting package, using the $\chi^{2}$ minimization technique. 
The Galactic column value used throughout is $N_{\rm H}=2.04\times 10^{20}$ cm$^{-2}$ \citep{gal}, and all line energies are quoted in the rest frame of the source.
\subsection{RGS}
\begin{figure*}
\centering
\vspace*{7.5cm}
\leavevmode
\includegraphics{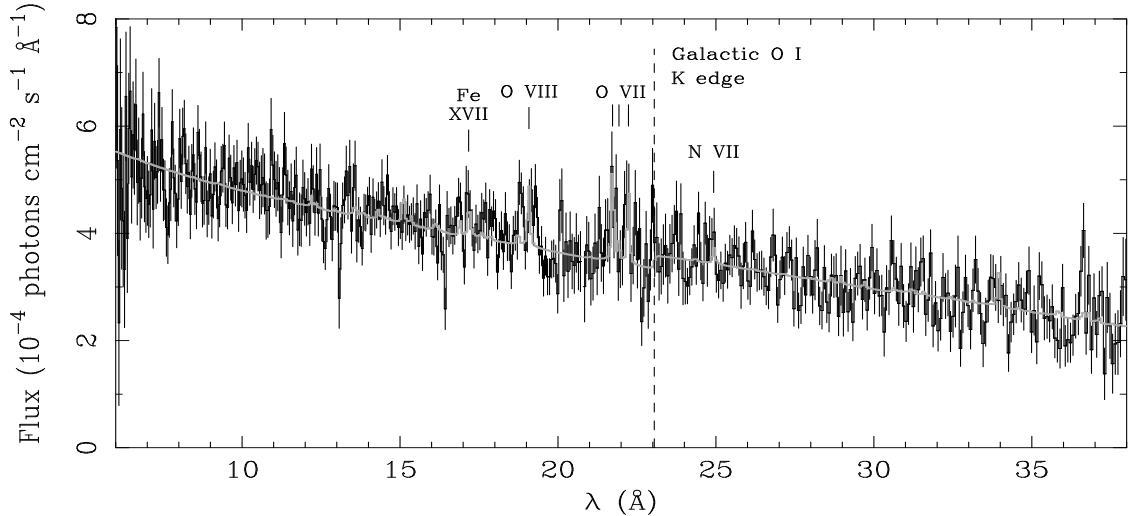}
\caption{The RGS spectrum of NGC~7213 in the source rest frame with best fitting model (grey line). The positions of prominent emission
lines are marked as well as the O~I K edge from the Galactic interstellar
medium.}
\label{fig:rgsspec}
\end{figure*}
\begin{table*}
\caption{Model fits to the RGS spectrum, with power law normalisations given in
photons keV$^{-1}$ cm$^{-2}$ s$^{-1}$ at 1 keV, $kT$ in keV, and {\small MEKAL}
normalisations ($= \int n_{e} n_{H} dV$) given in units of 
$10^{62}$~cm$^{-3}$. 
The Galactic column is included in the fits and all errors are quoted at the 90
per cent confidence level for 1 interesting parameter.}
\begin{tabular}{lccccc}
\hline
Model & $\Gamma$ & PL norm/10$^{-3}$ & $kT$ &{\small MEKAL} norm/10$^{-3}$& $\chi^{2}/\nu$ \\
&&&&&\\
PL & $1.77^{+0.02}_{-0.02}$&$5.88^{+0.06}_{-0.06}$ & - & - & 586/498\\
PL+{\small MEKAL} & $1.76^{+0.02}_{-0.02}$ &$5.85^{+0.06}_{-0.06}$& $0.18^{+0.03}_{-0.01}$ &$6.7^{+2.7}_{-2.4}$& 554/496\\
PL+2$\times${\small MEKAL} & $1.76^{+0.02}_{-0.02}$&$5.80^{+0.07}_{-0.06}$ &$0.18^{+0.02}_{-0.03}$,
$0.56^{+0.16}_{-0.22}$ &$6.7^{+2.7}_{-2.4}$, $3.1^{+1.6}_{-1.9}$ &546/494\\
\end{tabular}
\label{tab:rgsresults}
\end{table*}

The RGS spectrum of NGC~7213 is shown in Fig. \ref{fig:rgsspec}. The spectrum
is dominated by continuum emission, but shows emission lines, particularly
from O~VII and O~VIII; the most important of these are detailed in Table \ref{tab:lines}. 
No significant absorption lines nor broad absorption features such as
unresolved transition arrays (UTAs,
Behar, Sako \& Kahn 2001)
are observed. The features at 13.08~\AA~and 16.43~\AA\ in
Fig. \ref{fig:rgsspec}, which appear to be absorption lines, are actually low
signal to noise data points coincident with
chip-gaps in the first order spectra. 

\begin{figure}
\centering
\vspace{13.5cm}
\leavevmode
\includegraphics{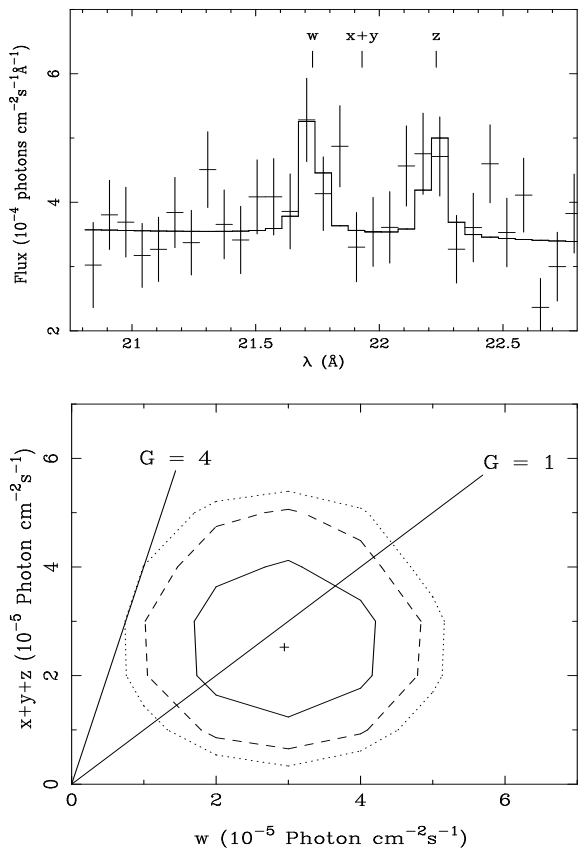}
\caption{Top panel: close up of the He-like O~VII triplet in the RGS spectrum
with best fitting power law plus 3-Gaussian model. Bottom panel: confidence
interval on the strength of the forbidden and intercombination lines (x+y+z)
against the resonance line (w). The solid, dashed and dotted contours
correspond to 68, 90 and 95 per cent respectively for two interesting parameters. 
The line $G=1$ indicates the ratio expected for
collisionally ionised plasma, while a photoionised plasma should lie to the
left of the $G=4$ line.}
\label{fig:ovii}
\end{figure}

The emission lines could come from gas which is either
photoionised or collisionally
ionised. The `$G$' ratio of the intercombination (x+y) and
forbidden (z) line strengths to the resonance (w) line strength in the He-like triplet of O~VII allows
the two cases to be discriminated (Porquet \& Dubau 2000), and therefore we
obtained the $G$ ratio by fitting the RGS spectrum in the 21-23 \AA\ region with a
power law and 3 emission lines, with energies fixed to the energies
of the O~VII triplet at the galaxies' redshift. The best fit is shown in the top panel of
Fig. \ref{fig:ovii} and the bottom panel shows the derived confidence 
interval on w and x+y+z. Collisionally ionised plasmas have $G\approx 1$,
fully consistent with that observed in NGC~7213
while photoionisation dominated plasmas have $G\geq 4$, which is 
excluded
at $>95$ per cent confidence. A photoionised plasma that does not lie along the line
of sight could have $G < 4$ if the resonance line is enhanced by
photoexcitation (Coup\'{e} \etal 2004). In this case, photoexcitation 
would also enhance the $3d-2p$ lines of Fe~XVII at $\sim$15 \AA\ relative to the
$3s-2p$ lines at $\sim$17 \AA, as is observed in NGC~1068 (Kinkhabwala \etal
2002) and Mrk~3 (Sako \etal 2000). However, this is not the case in NGC~7213, 
and so we conclude that the emission lines in the RGS spectrum are predominantly from collisionally ionised gas.

\begin{table}
\caption{Important emission lines in the RGS spectrum. Flux is given in photon cm$^{-2}$ s$^{-1}$.}
\label{tab:lines}
\begin{tabular}{llccc} \hline
$\lambda$               & ion  & Significance & EW               & Flux\\
(\AA)                   &      & ($\sigma$)   & (m\AA)
& /$10^{-6}$\\
&&&&\\ 
16.77$^{+0.10}_{-0.11}$ & Fe~XVII & 1.7 & 18$^{+16}_{-16}$ & 8$^{+7}_{-7}$\\
17.06$^{+0.12}_{-0.08}$ & Fe~XVII & 1.5 & 20$^{+20}_{-20}$ & 9$^{+9}_{-9}$\\
19.03$^{+0.26}_{-0.27}$ & O~VIII & 3.4 & 47$^{+18}_{-22}$ & 21$^{+8}_{-10}$\\
21.59$^{+0.05}_{-0.05}$ & O~VII  & 2.7 & 57$^{+32}_{-34}$ & 25$^{+14}_{-15}$\\
22.05$^{+0.06}_{-0.07}$ & O~VII  & 2.6 & 59$^{+40}_{-35}$ & 25$^{+17}_{-15}$\\
33.56$^{+0.39}_{-0.16}$ & C~VI   & 1.5 & 37$^{+39}_{-37}$ & 14$^{+15}_{-14}$\\
\end{tabular}
\end{table}

To model the RGS spectrum we began with a power law modified by the Galactic
absorption. The results for this and all the RGS spectral fits are given in
Table \ref{tab:rgsresults}. The best fitting photon index is $\Gamma=1.77\pm0.02$. However the $\chi^{2}/\nu$ is not good, and the model can be
rejected at $>99.5$ per cent confidence. To improve the fit we add a {\small MEKAL}
thermal plasma component, and obtain an acceptable fit with a best fit thermal
plasma temperature of $kT = 0.18^{+0.02}_{-0.01}$ keV. The model and data are shown in Fig. \ref{fig:rgsspec}. Addition of a second
thermal plasma component improves the fit only slightly ($\Delta \chi^{2} = 8$
for 2 extra parameters) with a best fit temperature of
$kT = 0.56^{+0.16}_{-0.22}$ keV for the extra component. 
The O~VII lines are reproduced well by the 0.18 keV thermal plasma
component, but there appears to be some emission adjacent to O~VIII
Ly$\alpha$ in excess of the model (either 1- or 2-temperature plasma)
prediction. This might indicate that the higher temperature component is 
physically
extended, a component of the host galaxy bulge rather than of the active
nucleus; alternatively, the emission could be broadened by doppler motions if
it originates within the broad line region of nucleus. Unfortunately, further
investigation of the O~VIII emission line profile is precluded by the
relatively low signal to noise ratio of the RGS spectrum.
We have also tried adding a blackbody soft excess component to the
model. This does not significantly improve the fit ($\Delta \chi^{2}$ = 6 for 2
extra parameters), and the $3 \sigma$ upper limit to the
contribution of a blackbody component is 11 per cent of the 0.3-2 keV flux.

\subsection{The EPIC soft X-ray spectrum}
Having found that a power law plus {\small MEKAL} component is a good fit to the RGS data, we began by fitting the same model to the 0.3 - 2 keV combined EPIC pn-MOS1 spectrum.
This gives $\chi^{2}/\nu=8865/33$. The fit is poor because the continuum slope
seen in EPIC
 is steeper than that of the model, with a data:model ratio of up to 1.3. If we
fix the slope and normalisation of the power law to their upper 90 per cent
confidence limits given by the fit to the RGS data, the fit is 
still not acceptable ($\chi^{2}/\nu=7349/33$).
Keeping the single {\small MEKAL} component as fitted to the RGS data, the power law slope must still be increased to 1.950$\pm$0.006 and normalisation to 6.56$\pm0.02 \times 10^{-3}$ photons cm$^{-2}$ s$^{-1}$ to obtain an acceptable fit to the 0.3 - 2 keV EPIC spectrum with $\chi^2$/$\nu$ = 60/35. Since the very low energies are less well calibrated for the EPIC cameras, we restrict the range to 0.5 - 2 keV. The continuum in this range can be adequately modelled by a single power law plus Galactic absorption ($\chi^2/\nu=345/208$). It is also well
fitted with a blackbody model ($\chi^2/\nu=305/208$), but trying instead a
multicolour accretion disk (diskbb) model we find the parameters of this model
cannot be constrained.
However, we cannot draw strong conclusions from this result since cross calibration of the three instruments is not accurately known. The MOS/pn cross correlation agrees to within 10 per cent from energies of 0.4 keV upwards and EPIC and RGS appear to agree in normalisation to $\pm$20 per cent, with the EPIC having a significantly steeper slope than the RGS in individual fits (Kirsch 2003). 
So the soft X-ray spectrum of this AGN shows some curvature, steepening towards lower energies. The soft-excess component noted by Bianchi \etal (2003) is a combination of this power-law-like excess and the {\small MEKAL} component. The existence of true soft-excess emission is, however, uncertain since the $\sim$14 per cent soft flux increase could be accounted for by uncertainties in the cross calibration of the EPIC and RGS instruments.

\begin{center}
\begin{table}
\caption{~Best fitting model to the 2 - 10 keV combined MOS1 and pn spectrum. The
Galactic column is included in all fits. All energies are given in the source restframe ($z$ =
0.005977): E in keV and EW in eV. The power law normalisation is given in photons cm$^{-2}$ s$^{-1}$. Errors are quoted at the 90 per cent confidence level for 1 interesting parameter.}
\begin{center}
\begin{tabular}{cc} \hline
$\Gamma$&1.73$^{+0.01}_{-0.01}$ \\
norm &5.65$^{+0.10}_{-0.10} \times 10^{-3}$\\
E$_{1}$ &6.41$^{+0.01}_{-0.02}$ \\
EW$_{1}$ &82$^{+10}_{-13}$\\
E$_{2}$ &6.66$^{+0.05}_{-0.05}$\\
EW$_{2}$ &24$^{+9}_{-11}$\\
E$_{3}$ &6.98$^{+0.07}_{-0.09}$\\
EW$_{3}$ &24$^{+10}_{-13}$\\
$\chi^{2}$/$\nu$&212/169 \\ 
\end{tabular} 
\end{center}
\label{tab:epicfits}
\end{table}
\end{center}

\subsection{The EPIC medium energy X-ray spectrum}
A power law plus Galactic absorption is clearly a poor fit to the 2 - 10 keV EPIC
data.
The residuals in the pn data between 6
and 7 keV were corrected by Bianchi \etal (2003) with the inclusion
of three emission lines, corresponding to Fe~I,
Fe~XXV and Fe~XXVI K$\alpha$. The availability of a
simultaneous \emph{BeppoSAX} PDS observation allowed these authors to
rule out the presence of a significant Compton reflection
component.
Combination of the EPIC pn and MOS1 data 
provides better statistics than pn alone, 
 allowing us to put tighter constraints on the parameters of spectral
features. We fit a power law with Galactic absorption, plus 3 Gaussian lines of
fixed narrow width ($\sigma$ = 0.001 eV) to the 2 - 10 keV combined pn-MOS1
spectrum. The best fit ($\chi^{2}$/$\nu$ = 212/169) has a power law photon
index of $\Gamma=1.73\pm0.01$, consistent with that found in the RGS soft X-ray
data.  
The centroid energies of the emission lines in the fit to the combined EPIC
data
are indeed consistent with iron fluorescence in low ionisation material, Fe XXV and Fe XXVI. These lines are well constrained and the results of the fit are given in Table \ref{tab:epicfits}.

\section{Discussion}
\subsection{Comparison with the Seyferts}
The 2 - 10 keV spectrum of NGC~7213 resembles a typical Seyfert
galaxy, which is dominated by a $\Gamma\sim1.7$ power law and a 6.4 keV 
Fe K$\alpha$ emission line. The 6.4 keV line in 
NGC~7213 is narrow (FWHM $<6820$ km~s$^{-1}$ at the 90 per cent confidence level), so it cannot
originate in the inner parts of an accretion disc. Bianchi \etal (2003) concluded that, 
given the apparent absence of a reflection
component in the \emph{BeppoSAX} data, the neutral Fe K$\alpha$ emission must
arise from Compton-thin material out of our line-of-sight, either in the 
form of a torus with a column density of 
$N_{\rm H}\sim 2 \times 10^{23}$ cm$^{-2}$ or the broad line
region. The resolution of the 
EPIC cameras does not allow the two to be distinguished.  
The 2 - 10 keV spectrum of NGC~7213 differs from that typically observed in 
Seyfert 1s
 in that it contains significant emission lines from Fe~XXV and
Fe~XXVI. 
These lines are not normally observed in the classical 
luminous Seyfert galaxies
(e.g. NGC~3783, Kaspi \etal 2002; NGC~5548, Pounds \etal 2003; NGC~7469,
Blustin \etal 2003), though they have been observed in some cases
(e.g. NGC~5506, Bianchi \etal 2004). In contrast, these lines appear to
dominate the Fe~K$\alpha$ emission in the nearby
LINER M\,81 (Page \etal 2004). 
The origin of the Fe~XXV and
Fe~XXVI lines in M\,81 is unclear (see Page \etal 2004 for a discussion), and
this is also true for NGC~7213. They may be produced by photoionisation of
Compton-thin material by the nuclear X-ray source (Bianchi \etal 2004), or may
be collisionally ionised like the soft X-ray thermal plasma. Future
observations of the Fe~K region at higher spectral resolution, for example
with the forthcoming {\em Astro-E2} mission, might answer this question by
allowing the application of line diagnostics to the Fe~XXV triplet.

In the soft X-ray regime, NGC~7213 departs further from a Seyfert-like 
spectrum.
The ionised gas which is seen in the soft X-ray spectra of Seyfert galaxies is
usually found to be photoionised, whether it is seen in 
absorption or in emission (e.g. IRAS 13349+2438, Sako \etal 2001; NGC 3783, 
Kaspi \etal 2002; NGC 1068, Kinkhabwala \etal 2002). 
However, in the RGS spectrum of NGC~7213 we do not find any evidence for 
photoionised absorption, and the emission lines 
are collisionally ionised rather than photoionised (Fig. \ref{fig:ovii}).
Amongst AGN, collisionally ionised thermal plasma at soft X-ray temperatures,
as observed in NGC~7213, may be a property specific to LINERs. 
In a study of 21 LINERs observed with \asca, most were found to be well fitted with a hard X-ray power 
law ($\Gamma \sim 1.8$) plus a soft X-ray thermal plasma component 
with $kT < 1$ keV \citep{terash}. 
While this provides strong evidence for the presence of a soft X-ray emission 
line component in LINERs, the resolution of \asca\ 
was not good enough to permit
diagnostics which could distinguish between collisionally ionised and
photoionised plasma. However, the very nearby LINER 
galaxy M\,81, for which there is a long-duration RGS spectrum, does contain a
significant component of soft X-ray collisionally-ionised thermal 
plasma \citep{M81}.

The \xmmn\ spectrum of NGC~7213 also deviates from that of a typical Seyfert in that it has
little or no blackbody-like soft X-ray excess emission. This component is found
in the X-ray 
spectra of most Seyferts (Pounds \& Reeves 2002; Turner \& Pounds 1989), the strongest being seen in the Narrow-line Seyfert 1 galaxies
(NLS1s).
The origin of the soft X-ray excess is unknown, but is often interpreted as the high energy tail of the thermal emission from an optically thick
accretion disc \citep{exosat}, particularly in NLS1s (e.g. RE 1034+39,
Pounds, Done \& Osborne 1995). 

\begin{figure}
\centering
\vspace*{7.5cm}
\leavevmode
\includegraphics{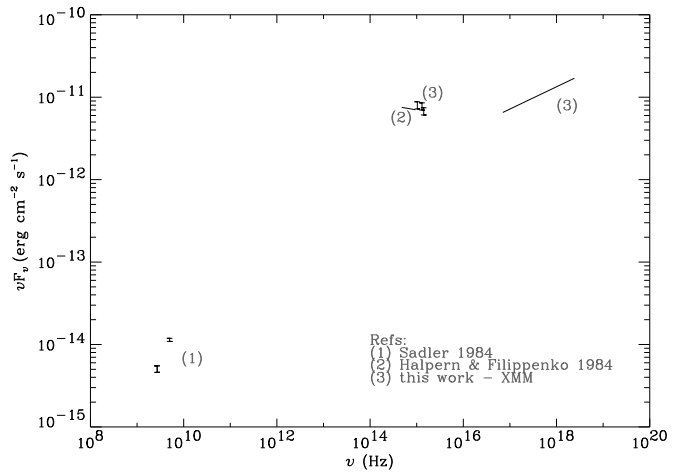}
\caption{The AGN radio to X-ray continuum spectral energy distribution for NGC~7213. Note the data are not simultaneous measurements.}
\label{sed}
\end{figure}   
\subsection{Evidence for an accretion disc}
Many Seyfert galaxies show strong evidence in X-rays for an accretion disc 
surrounding the black hole. The main indicators are a soft-excess, reflection,
and broad Fe K$\alpha$ line 
emission, all of which come from the inner parts of the accretion disc. 
None of these indicators are present in the \xmmn\ spectra of NGC~7213.

The UV bump in NGC~7213 is either absent or extremely weak compared to other
Seyferts \citep{UV}. The optical continuum emission of NGC~7213 comprises
both non thermal emission from the AGN and stellar radiation (HF84). A
decomposition of the spectrum, assuming a power law for the AGN contribution
following $F_\nu \propto \nu^{-1.1}$, shows that more than 50 per cent of the total
continuum emission at 3300 \AA\ can be non-stellar. The actual flux level is
uncertain by $\pm$30 per cent and also depends on the visual extinction
adopted. HF84 argue that the optical extinction is much higher than the
Galactic value. 
However the \xmmn\ data show no evidence for any intrinsic
absorption of the X-ray flux. Therefore we have adopted the AGN power law
component of the optical spectrum, deconvolved from the stellar contribution by HF84, and assumed only Galactic
absorption (A$_{V}$ = 0.05). The nuclear spectral energy distribution of NGC~7213 is shown in 
Fig. \ref{sed}. The ultraviolet data from the OM are in good agreement with
an extrapolation of the optical power law to shorter wavelengths.
There is no evidence for an 
optical/UV bump and consequently the AGN bolometric luminosity does not
appear to be dominated by emission from an optically thick,
geometrically thin accretion disc.

Assuming that NGC~7213 does contain an accretion disc, there are two
possible explanations for the lack of UV emission. Either the disc does not
extend close enough to the black hole to emit significantly in the 
optical/ultraviolet, or the disc extends close to the black hole but has
such a low mass transfer rate that it contributes little to the overall
emission.

Starting with the truncated inner-disc explanation, it is known that at low
accretion rates, stable accretion flows can exist in which the inner parts
of the disc take the form of an optically-thin, tenuous, hot corona. Because the inner, hot corona is
a much less efficient radiator than a thin disc, much of the energy may be
advected into the black hole rather than being radiated. This is known generally as an
advection-dominated accretion flow (ADAF, Narayan \& Yi 1995), and several derivatives of this model have since been developed to include, for example, more realistic geometries (e.g. CDAF model, Narayan, Igumenshchev \& Abramowicz 2000) and/or outflows (e.g. ADIOS model, Blandford \& Begelman 1999), all of which require a truncated disc.
A truncated inner disc plus ADAF is the configuration proposed to explain
the spectra of black hole X-ray binaries 
(Esin \etal 1998; Esin, McClintock \& Narayan 1997; 
Narayan, Garcia \& McClintock 1997). The quiescent phase of the soft X-ray
transients V404 Cyg and A0620-00 can be explained with the ADAF model, where
the inner radius of the accretion disc lies at $\sim10^{4}$ Schwarzschild
radii \citep{NBM97}. To explain optical
and UV observations of M\,81, Quataert
\etal (1999) invoke an ADAF within the truncation radius of the disc.
An ADAF model has also been successfully applied to describe the
multiwavelength emission from the Galactic Centre \citep{ADAFGC}. In ADAF-type models the soft X-ray and UV signatures of thermal disc emission are weak or
absent because the geometrically thin, optically thick part of the disc does
not extend close enough to the black hole. However, such accretion flows are also expected to produce outflows not yet observed in NGC~7213. 

An alternative to a truncated disc is a disc which extends to 
the inner regions but which is in a `low state'; this can also account for the 
weak or absent disc emission from NGC~7213.
Thermal-viscous ionisation instabilities, which can develop in a partial
ionisation zone of an AGN accretion disc, can cause 
the disc to oscillate between high (bright) states and low (faint) states
\citep{aneta}. 
The disc spends most of the time in the low state with low luminosity,
peaking in the IR and emitting negligible optical and UV radiation. This is 
analogous to the accretion disc instability which drives the dwarf nova outbursts in cataclysmic variables (Meyer \& Meyer-Hofmeister 1984). 
However, although a low-state disc can account for the lack of primary disc
emission, it ought to constitute an X-ray reflector subtending a substantial 
solid angle to the X-ray source. 

\subsection{Accretion rate}
Both the ADAF and the `low-state' disc models imply a significantly sub-Eddington accretion rate (e.g. a few percent of the Eddington rate or less for the ADAF model, Narayan \& Yi 1995). 
To investigate the accretion rate in NGC~7213
we have estimated the bolometric luminosity of the active nucleus by 
integrating the radio to X-ray spectral energy distribution.
For this we have used the radio measurements from Sadler (1984), the 
power law component in the
optical, deconvolved from the stellar contribution by Halpern \& Filippenko
(1984), the UV flux measurements from the \xmmn\ OM and the \xmmn\ X-ray
power law continuum. The bolometric luminosity thus obtained is
9$\times 10^{42}$ erg s$^{-1}$.
Previous estimates of the bolometric
luminosity are significantly higher than our determination: $L_{\rm bol}=10^{44.3}$
erg s$^{-1}$ \citep{bhmass} and $L_{\rm bol} = 10^{44.1}$ erg s$^{-1}$
\citep{mass2}. However, these values include the host galaxy as well as the
AGN, so will overestimate the AGN luminosity by a considerable amount.
The black hole mass has been estimated from the stellar velocity
dispersion by Nelson \& Whittle (1995), who obtained
$M_{\rm BH}=10^{8.0} \msun$.
Such a high black hole mass is consistent with
the observed lack of significant X-ray variability during the 46\,ks \xmmn\
observation.
Combining the bolometric luminosity and mass estimates we find that the 
luminosity of NGC~7213 is approximately
7$\times 10^{-4}~L_{\rm Edd}$. Generally,
Seyfert luminosities lie in the range 0.001-1 $L_{\rm Edd}$ (Wandel
1999). Padovani \& Rafanelli (1988) found an
average luminosity of $\sim 0.2~L_{\rm Edd}$ for the 34 local Seyfert 1 galaxies within their sample of
AGN, although a much lower luminosity of $\sim 0.005~L_{\rm Edd}$ is inferred for
local type 1 Seyfert galaxies from studies of their X-ray luminosity function
(Page 2001).
In any case, our luminosity estimate for NGC~7213 lies at or below the low
end of the distribution of Seyfert luminosities. M\,81, the nearest known LINER, radiates at 4.2$\times 10^{-4}$ times its Eddington rate \citep{Ho1999}, 
and a rate of
10$^{-4}~L_{\rm Edd}$ was determined for the LINER NGC~4203
\citep{4203}.
 The low
luminosity AGN which are found in LINER galaxies may be typified by
lower accretion rates than Seyferts, assuming approximately the same efficiency of mass accretion.

\subsection{Comparison with M\,81: the Seyfert--LINER connection}
An interesting comparison can be made between the X-ray spectra of NGC~7213 and
M\,81, since M\,81 is 
the nearest LINER galaxy and has been studied in detail using \xmmn\
(Page \etal 2003; Page \etal 2004).
The broad-band X-ray spectra of these two galaxies, at first glance, look
remarkably similar, but whilst the continua are comparable we find
substantial differences in the emission line parameters.
The soft X-ray emission lines are less prominent in NGC~7213 than in M\,81,
and the emission from collisionally ionised thermal plasma components in NGC~7213 comprises a
much smaller
fraction of the total 0.3 - 10 keV flux than in M\,81 (0.3 per cent, compared
with 8.7 per cent in M\,81). Emission lines arising in collisionally ionised plasma are not normally observed
in Seyfert galaxies. 
On the other hand the Fe~I K$\alpha$ line at 6.4
keV is commonly found in Seyferts \citep{nandra},
and the equivalent width of this line is
about twice as large in NGC~7213 as it is in M\,81.
Therefore, although NGC~7213 contains the soft X-ray emission lines and
Fe~K$\alpha$ lines of Fe~XXV and Fe~XXVI that are not usually observed in
Seyfert galaxies but are seen in the LINER galaxy M\,81, the relative
weakness of these lines and the strength of Fe~I~K$\alpha$ in NGC~7213 make
the X-ray spectrum of NGC~7213 much more Seyfert-like than that of M\,81.
Therefore NGC 7213 appears to bridge the gap
between `normal' Seyfert galaxies and LINER galaxies such as M\,81.

It is likely then, that there is a continuous distribution of galaxy nuclei
between the LINERs and `normal' Seyfert nuclei, over which the X-ray spectral 
features characteristic of Seyferts such as the neutral Fe K$\alpha$ line, 
become
successively more prominent, while the features characteristic of LINERs such
as soft X-ray emission lines diminish in significance. 
Accretion rate onto the black hole with respect to the Eddington rate is 
likely to be the overriding factor,
with LINER galaxies accreting at much lower rates than 
Seyfert galaxies (Ho, Filippenko \& Sargent 2003) and containing truncated discs.
In fact, if we look at the observational properties of the Galactic Centre
we find that it may also fit into this continuous distribution, at the
opposite end of the scale to the Seyferts. The Galactic Centre contains a
low-mass black hole with an extremely low accretion rate
($1.8\times10^{-6} \le \dot{M}/\dot{M}_{\rm Edd} \le 1.5\times10^{-3}$, Falcke \& Biermann 1999; Quataert, Narayan \& Reid 1999) and with a luminosity of $<1\times 10^{-6}~L_{\rm Edd}$.
The emission from this region comes predominantly from thermal plasmas
with strong soft X-ray emission (Baganoff \etal 2003). At higher energies Fe
K$\alpha$ emission is observed (Tanaka \etal 2000), the strongest line being
at 6.7 keV. Therefore, the Galactic Centre begins the sequence: its
characteristics being a very low $L/L_{\rm Edd}$ ratio, dominance of thermal plasma
emission and strong, highly ionised iron emission. 

The lack of reflection in NGC~7213 may provide a clue to the explanation of
the low luminosities in LINER galaxies. Although there is a significant
6.4~keV line, the lack of reflection implies that this must arise in
Compton-thin material. Thus the central region of NGC~7213 appears to be
deficient in the dense, cool material that usually gives rise to reflection
in Seyferts. In M\,81, which has an even lower luminosity than NGC~7213,
the 6.4~keV line is only half as strong as it is in NGC~7213, implying that
the central regions of M\,81 are even more lacking in cool material than
NGC~7213. That LINERs have gas-poor central regions relative to Seyferts has
also been proposed by Ho \etal (2003) on the basis of their
optical emission line properties. It therefore appears that the low luminosities and
accretion rates of LINER-AGN are a consequence of the shortage of material in
their central regions. In this case LINERs are simply fuel-starved AGN, and
could represent the weak, but not yet silent remnants that have evolved from
a previous generation of Seyferts and QSOs.

\section{Conclusions}
We present an \xmmn\ observation of the LINER galaxy NGC~7213 with the EPIC and RGS X-ray instruments and UV photometry with the OM.
The X-ray spectrum is typical of a Seyfert galaxy, although with a much lower luminosity ($L_{2-10 keV}=1.7\times 10^{42}$ erg s$^{-1}$). A collisionally ionised thermal plasma of temperature 0.18$^{+0.03}_{-0.01}$ keV is
required in the RGS soft X-ray region where several low ionisation emission
lines are detected, and we find no evidence for a significant blackbody-like soft-excess component. We suggest that this low ionisation thermal plasma may
be a LINER characteristic.
We constrain the equivalent widths of Fe I, XXV and XXVI K$\alpha$ emission lines to 82$^{+10}_{-13}$, 24$^{+9}_{-11}$ and 24$^{+10}_{-13}$ eV respectively.

The nuclear radiation at X-ray
wavelengths in NGC~7213 is substantially more Seyfert-like than the AGN component in the 
nearby LINER galaxy M\,81, meaning NGC 7213 has the properties of a galaxy
somewhere in between a typical Seyfert and a LINER galaxy. This
supports the notion of continuity between the LINER and Seyfert classes,
dictated by the luminosity of the central AGN and its accretion rate,
both increasing from LINERs to Seyferts. 
In turn, the mass accretion rates depend critically on the 
amount of material present in the central regions surrounding the AGN.
This continuous distribution can
be extended to include the Galactic Centre, which shares some properties
with LINER galaxies.

\section{Acknowledgments}
This work is based on observations obtained with \xmmn, an ESA science
 mission with instruments and contributions directly funded by ESA Member
 States and the USA (NASA). 
RLCS acknowledges support from a PPARC studentship.

\bsp

\label{lastpage}

\end{document}